\documentclass[english,english,pra,english,preprint,amsmath,amssymb,aps,longbibliography,showkeys, titlepage]{revtex4-2}
\pdfoutput=1 
\usepackage{lmodern}
\usepackage[T1]{fontenc}
\usepackage[latin9]{inputenc}
\usepackage{amsmath}
\usepackage{amsthm}
\usepackage{amssymb}
\usepackage{graphicx}
\usepackage{wasysym}

\makeatletter
\usepackage[T1]{fontenc}
\usepackage{amsmath}
\usepackage{amsthm}
\usepackage{amssymb}
\usepackage{graphicx}

\usepackage{graphicx}
\usepackage{float}
\usepackage{xcolor}
\usepackage[hypertexnames=false]{hyperref}
\hypersetup{
	breaklinks = true,
    colorlinks = true,
    citecolor = {blue},
	urlcolor = {blue},
	linkcolor = {blue}
}

\makeatother

\usepackage{babel}
\begin{document}
\title{Advanced fuel fusion, phase space engineering, and structure-preserving
geometric algorithms}
\author{Hong Qin}
\email{hongqin@princeton.edu }

\affiliation{Princeton Plasma Physics Laboratory, Princeton University, Princeton,
NJ 08540}
\affiliation{Department of Astrophysical Sciences, Princeton University, Princeton,
NJ 08540}
\begin{abstract}
Non-thermal advanced fuel fusion trades the requirement of a large
amount of recirculating tritium in the system for that of large recirculating
power. Phase space engineering technologies utilizing externally injected
electromagnetic fields can be applied to meet the challenge of maintaining
non-thermal particle distributions at a reasonable cost. The physical
processes of the phase space engineering are studied from a theoretical
and algorithmic perspective. It is emphasized that the operational
space of phase space engineering is limited by the underpinning symplectic
dynamics of charged particles. The phase space incompressibility according
to the Liouville theorem is just one of many constraints, and Gromov's
non-squeezing theorem determines the minimum footprints of the charged
particles on every conjugate phase space plane. In this sense and
level of sophistication, the mathematical abstraction of phase space
engineering is symplectic topology. To simulate the processes of phase
space engineering, such as the Maxwell demon and electromagnetic energy
extraction, and to accurately calculate the minimum footprints of
charged particles, recently developed structure-preserving geometric
algorithms can be used. The family of algorithms conserves exactly,
on discretized spacetime, symplecticity and thus incompressibility,
non-squeezability, and symplectic capacities. The algorithms apply
to the dynamics of charged particles under the influence of external
electromagnetic fields as well as the charged particle-electromagnetic
field system governed by the Vlasov-Maxwell equations. 
\end{abstract}
\maketitle

\section{Introduction}

Among all light-ion nuclear fusion reactions that are potentially
suitable for energy production, deuterium-tritium (D-T) fusion is
advantageous in terms of having the largest cross-section and lowest
energy for peaked cross-section. D-T fusion reactions can be sustained
in a confined, thermalized plasma consisting of deuterium and tritium
ions at 10KeV temperature without significant radiation loss. This
is desirable since fusion cross-sections are always much smaller than
those of Coulomb scatterings between charged particles, and the fusion
plasma thermalizes quickly before fusion unless powerful external
electromagnetic fields are applied to maintain the non-thermal particle
distributions. The cross-sections for advanced fuel fusion reactions
\citep{Bussard1994,Rostoker1997,Nevins1998,Volosov2006}, such as
the deuterium-helium-3 (D-He3) fusion and proton-boron-11 (P-B11)
fusion, peak at much higher energies. These fusion reactions are difficult
to sustain in thermalized plasmas because of the significantly increased
radiation loss of energy at higher temperatures \citep{Nevins1998,Son2006}.
From the perspective of reducing transport due to collisions and turbulence,
D-T fusion is also favorable because of the lower temperature required. 

\begin{figure}[ht]
\centering \includegraphics[height=8cm]{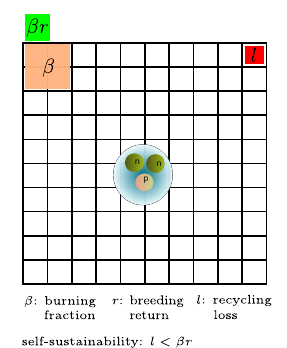} \caption{Tritium self-sustainability is still an engineering challenge. For
the tritium burning fraction and breeding return currently achievable,
there is a large amount of recirculating tritium in the system, and
the required tritium recycling and recovery rate is in the range of
99.9\%. }
\label{tritium}
\end{figure}

The disadvantages of D-T fusion are also prominent. First, the fusion
energy is carried by 14MeV neutrons, which damage the first wall and
other necessary plasma-facing hardware, such as RF antennas. Moreover,
magnetic confinement devices such as tokamaks or stellarators will
require massive superconducting magnets, which will need to be fully
shielded from these neutrons. Much effort has been directed toward
developing materials that can resist the energetic fusion neutrons.
There is another difficulty. D-T fusion power plants are believed
to be clean and have an unlimited supply of fuel. This is certainly
true for deuterium. But for tritium, fuel self-sustainability is still
an engineering challenge \citep{Sawan2006,Abdou2020}. Assuming initial
tritium inventory is available through other venues, tritium breeding
using fusion neutrons and lithium-6 is the currently envisioned solution
for tritium self-sustainability. However, this requires a near-perfect
tritium recycling and recovery rate, given the tritium burning fraction
and tritium breeding ratio currently achievable. For an analogy to
draw below, let $\beta$ denote the tritium burning fraction in one
particle confinement time, $r$ the tritium breeding return (1+$r$
is the tritium breeding ratio), and $l$ the tritium recycling and
recovery loss rate. For simplicity, we use the same loss rate for
the tritium recycling from divertors/first walls and tritium recovery
from the breeder blankets. Tritium self-sustainability means $\beta+(1-\beta)l<\beta(1+r)(1-l)$,
which, when $\beta$ and $l$ are small, simplifies to 
\begin{equation}
l<\beta r.
\end{equation}
For the typical values of $\beta=1\%$ and $r=10\%$ \citep{Sawan2006,Abdou2020},
the tritium recycling and recovery loss rate $l$ must be less than
$0.1\%.$ This is a stringent requirement imposed by the large amount
of recirculating tritium in the system associated with the low tritium
burning ratio (see Fig.\,\ref{tritium}). Even if a near-perfect
recycling and recovery rate is achievable, it translates to expansive
capital and operational costs. 

For commercial fusion energy production, advanced fuel fusion using
D-He3 or P-B11 has been considered as a possible alternative with
certain advantages \citep{Bussard1994,Rostoker1993,Rostoker1997,Nevins1998,Mazzucato2023,Lerner2023,Kong2023,wei2023,liu2024enns}.
In addition to being tritium-free, D-He3 and P-B11 reactions are also
almost aneutronic, removing the need to develop neutron-resisting
materials. In exchange for these two major advantages, advanced fuel
fusion needs to be operated in a non-thermal setting, for the reason
discussed above, and the challenge is how to maintain the non-thermal
particle distributions and the associated recirculating power at an
affordable cost \citep{Bussard1994,Rostoker1993,Rider1997,Rostoker1997,Nevins1998a}.
In this sense, advanced fuel fusion trades the requirement of a large
amount recirculating tritium for that of large recirculating power.
It is reasonable to expect that recirculating power is easier to generate
and maintain in comparison. To meet the challenge, innovative phase
space engineering using electromagnetic fields can be applied \citep{Ochs2021,Kolmes2022a,Ochs2022,Munirov2023,Mlodik2023,Ochs2024}.
For charged particles in ionized plasmas, injecting high-frequency
electromagnetic fields seems to be the only way to manipulate phase
space distributions faster than the collisional thermalization, and
such methods have been extensively investigated for achieving hot
ion modes via $\alpha$-channeling in D-T fusion \citep{Fisch1992,Fisch1995a,Fisch1995,Herrmann1997,Ochs2015}.
In the present study, we analyze the physics of phase space engineering
from a theoretical and algorithmic perspective. In particular, we
emphasize the phase space constraints imposed by the Hamiltonian nature
of the dynamics and discuss structure-preserving geometric algorithms
that are indispensable in the study and design of phase space engineering
techniques. In Sec.\,\ref{sec:pse}, we discuss the need for phase
space engineering for non-thermal advanced fuel fusion. Symplectic
dynamics of charged particles, the underpinning physics of phase space
engineering, and the associated phase space constraints are analyzed
in Sec.\,\ref{sec:sd}. Structure-preserving geometric algorithms
suitable for phase space engineering are presented in Sec.\,\ref{sec:spa}. 

\section{Phase space engineering for non-thermal advanced fuel fusion \label{sec:pse}}

Even for the thermal D-T fusion, the fusion plasma is not fully thermalized.
Phase space engineering has been utilized to heat plasmas \citep{Stix1992,Dodin2022},
suppress instabilities \citep{Reiman1983,Yoshioka1984,Haye2006,Reiman2018},
drive current \citep{Fisch1987}, and channel the energy of $\alpha$-particles
for hot ion modes \citep{Fisch1992,Fisch1995a,Fisch1995,Herrmann1997,Ochs2015}.
In particle accelerators, phase space engineering is routinely used
to control the properties of charged particle beams \citep{Courant1958,Chao93-all,Wangler98-all,Davidson01-all}. 

\begin{figure}[ht]
\centering \includegraphics[width=8cm]{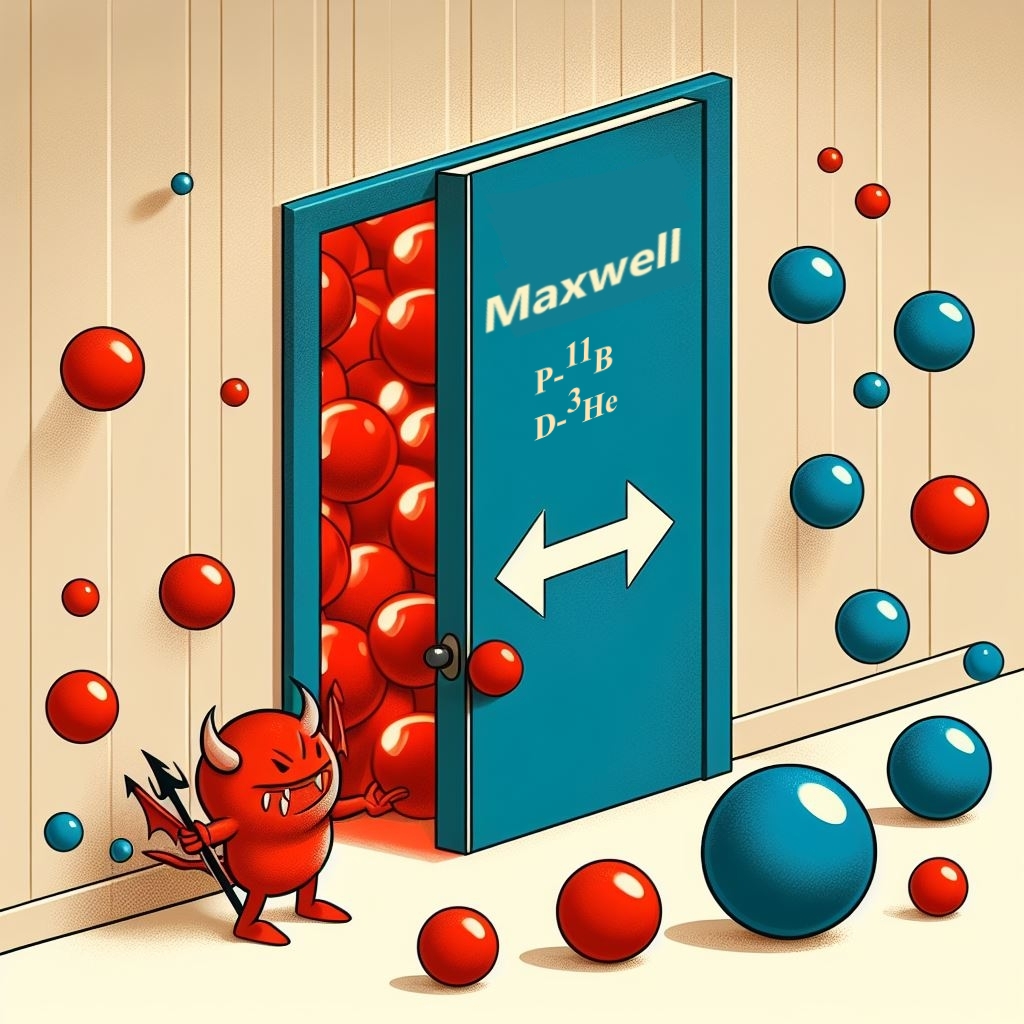} \caption{A Maxwell demon guarding a one-way door. The blue balls inside (representing
backward-moving particles) are bounced back at the door, and the red
balls outside (representing forward-moving particles) are allowed
in. }
\label{Maxwell}
\end{figure}

For non-thermal advanced fuel fusion, large recirculating power exists
in the system to maintain the non-thermal distributions of particles,
which does not necessarily entail significant energy loss if enough
attention is paid to preserve the power flow. An example is the energy-recovering
accelerator technology \citep{Schliessmann2023}. If the success of
D-T fusion energy is only possible with lossless tritium recirculation
in the system, non-thermal advanced fuel fusion depends on phase space
engineering that can sustain efficient recirculating power. Let's
demonstrate conceptually how an RF field-driven Maxwell demon \citep{Szilard1929,Brillouin1951,Landauer1961,Bennett1982,Zurek1989,Fisch1992,Landauer1996,Fisch2003,Dodin2004}
can be utilized to maintain the directed energy of a fusing ion beam.
In the co-moving frame of the beam, a Maxwell demon is implemented
as a one-way door that allows ions with forward velocity to pass through
but reflects ions with backward velocity, see Fig.\,\ref{Maxwell}.
The one-way door guarded by the demon can also be called a one-way
wall. It converts the thermal energy into the directed kinetic energy
of the beam. In this process, the entropy of the ion beam is decreased,
which is associated with certain energy costs and increased entropy
of the entire system including the external hardware generating the
RF field. The technical challenge is to minimize the cost within the
operational space allowed by physics. 

\begin{figure}[ht]
\centering \includegraphics[width=8cm]{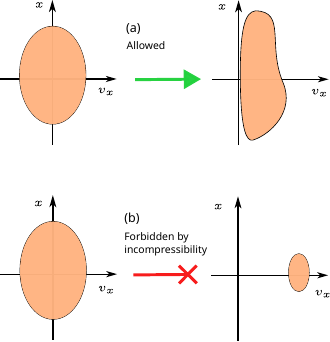} \caption{Not all phase space manipulations by a Maxwell demon driven by electromagnetic
fields are possible. The phase space volume of the ions needs to be
conserved according to the Liouville theorem. If the operation is
confined in the $(x,v_{x})$ plane, incompressibility rules out the
possibility of forming a cold, focused, energetic beam as shown in
(b), which would be ideal for fusion reactions. }
\label{AllowedForbiddenMD}
\end{figure}

Not all phase space manipulations are possible if the Maxwell demon
was driven by electromagnetic fields. For example, the phase space
volume of the ions needs to be conserved according to the Liouville
theorem, which is known as phase-space incompressibility (see Sec.
\ref{sec:sd}). When the backward-moving ions are pushed forward by
the demon, they have to occupy the same volume in the left part of
the phase space as shown in Fig.\,\ref{AllowedForbiddenMD}(a). If
the operation is confined in the $(x,v_{x})$ plane, incompressibility
rules out the possibility of forming a cold, focused, energetic beam
as shown in Fig.\ref{AllowedForbiddenMD}(b), which would be ideal
for fusion reactions. 

One of the advantages of aneutronic fusion is that the fusion energy
is released in the form of kinetic energy carried by charged particles,
and the fusion energy can be extracted directly and efficiently as
electromagnetic energy. However, because the kinetic energy of the
fusing ions is much smaller than the fusion energy released, the released
energy is largely thermalized and occupies a certain phase space volume
that needs to be conserved during the direct energy extraction via
electromagnetic channels. For a given distribution of fusion products,
e.g., $\alpha$-particles in the P-B11 fusion, what is the maximum
fusion energy that can be extracted electromagnetically \citep{Helander2017,Kolmes2020a,Helander2020,Kolmes2022}?
Or equivalently, what is the ground state of the system via electromagnetic
channels? Because of the phase-space incompressibility, the ground
state energy is not zero. 

\begin{figure}[ht]
\centering \includegraphics[width=8cm]{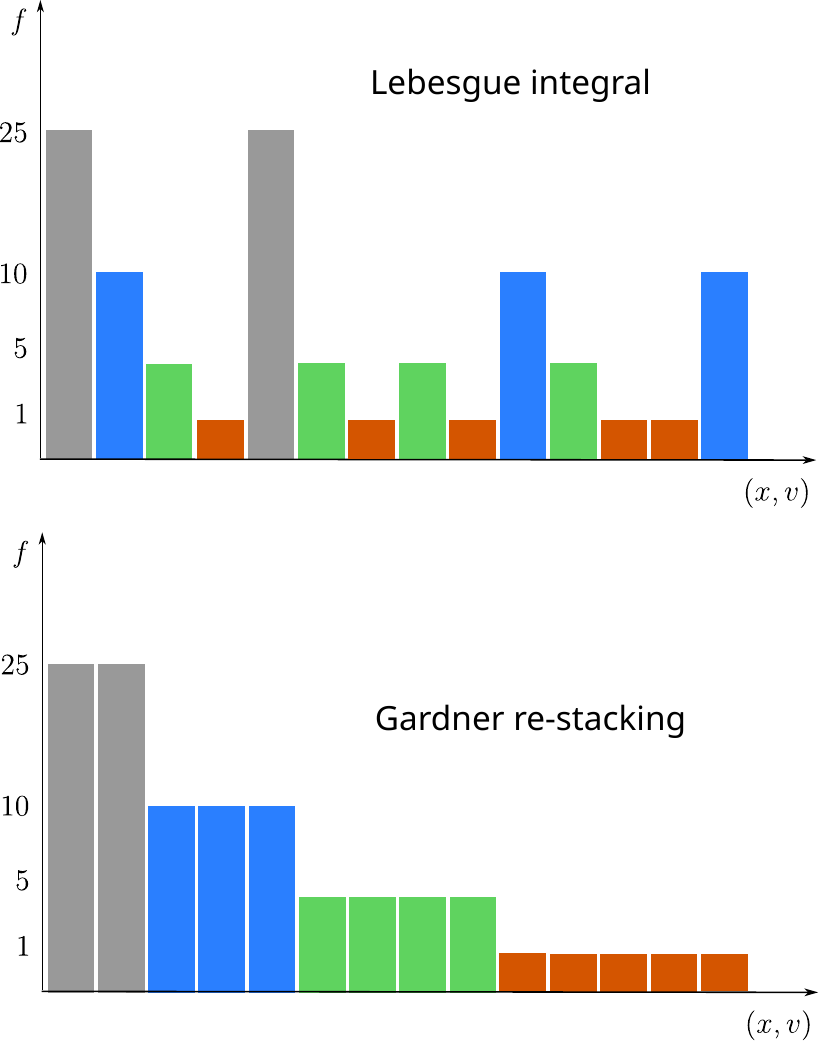} \caption{Gardner ground state is reached by ``doing a Lebesgue integral''. }
\label{Gardner}
\end{figure}

Gardner studied the ground state of a phase space distribution accessible
by all possible volume-preserving maps of phase space \citep{Gardner1963}.
Let $f_{0}(\boldsymbol{x},\boldsymbol{v})$ denote the initial distribution
and $\phi(\boldsymbol{x})$ an external potential. Gardner showed
that the ground state accessible via volume-preserving maps is $f_{1}(\phi(\boldsymbol{x})+mv^{2}/2)$,
where $f_{1}$ is a function of one scalar argument satisfying: (1)
$f$ is monotonically decreasing and (2) for any $\xi>0$, the phase
space volume of the region where $f_{1}(\phi(\boldsymbol{x})+mv^{2}/2)>\xi$
is equal to the phase space volume of the region where $f_{0}(\boldsymbol{x},\boldsymbol{v})>\xi$.
The function $f_{1}$ is uniquely determined. Let's call this ground
state accessible via volume-preserving maps Gardner ground state.
Intuitively, the Gardner ground state is reached by ``doing a Lebesgue
integral'', which can be explained as follows. Suppose a coin purse
dropped and the coins inside scattered on the floor. There are two
ways to pick up the coins and count your assets. You can pick up all
the coins on the first square foot of the floor, then all the coins
on the second square foot, and so on. This is the Riemann integral.
Alternatively, you can pick up all the quarters first, then all the
dimes, then all the nickels, and finally all the pennies. This is
the Lebesgue integral, and in Gardner's terminology, $\xi=25\cent,10\cent,5\cent,$
and $1\cent$. Gardner's construction was later interpreted as Gardner
re-stacking algorithm \citep{Dodin2005,Kolmes2020}. The algorithm
moves all the quarters to the center of the floor without changing
the size of their footprints, then all the dimes to the region next
to the quarters, and so on. See Fig.\,\ref{Gardner}. 

It turns out that incompressibility, or phase space volume conservation,
is just one of many constraints in phase space. Another important
constraint is known as non-squeezability, which is a consequence of
the symplectic nature of the Hamiltonian dynamics of charged particles
under the influence of electromagnetic fields. To design effective
phase space engineering schemes, it is crucial to understand these
dynamic constraints and adopt structure-preserving simulation algorithms
that preserve these constraints. As we will see, the Gardner ground
state may not be accessible by the Hamiltonian dynamics of charged
particles under the influence of externally injected or self-consistently
generated electromagnetic fields. 

\section{Symplectic dynamics of charged particles --- the physics of phase
space engineering\label{sec:sd}}

In this section, we briefly describe the constraints of phase space
dynamics as imposed by the law of physics and the relevance to phase
space engineering. The dynamics of charged particles under the influence
of an external or self-consistently generated electromagnetic field
is governed by Hamilton's equation, 
\begin{align}
\dot{x}^{i} & =\frac{\partial H}{\partial p_{i}},\,\,\,\dot{p}_{i}=-\frac{\partial H}{\partial x^{i}},\,\,\,i=1,2,3,\label{eq:HE}
\end{align}
where $H=H(x^{i},p_{i},t)$ is the Hamiltonian function. For a general
electromagnetic field on spacetime specified by 4-potential $(\phi(\boldsymbol{x},t),\boldsymbol{A}(\boldsymbol{x},t)),$
\begin{align}
\boldsymbol{p} & =m\dot{\boldsymbol{x}}+\frac{q}{c}\boldsymbol{A}(\boldsymbol{x},t),\label{eq:p}\\
H & =\frac{1}{2m}\left(\boldsymbol{p}-\frac{q}{c}\boldsymbol{A}(\boldsymbol{x},t)\right)^{2}+q\phi(\boldsymbol{x},t).\label{eq:H}
\end{align}
Here, we have taken the non-relativistic limit of $H$ for simplified
presentation. The dynamics according to Eqs.\,(\ref{eq:HE})-(\ref{eq:H})
is given by a solution map 
\begin{equation}
\varphi_{t}:\boldsymbol{z}(0)=\left(\boldsymbol{x}(0),\boldsymbol{p}(0)\right)\mapsto\boldsymbol{z}(t)=\left(\boldsymbol{x}(t),\boldsymbol{p}(t)\right).
\end{equation}
It is easy to verify that the solution map $\varphi_{t}$ is symplectic,
which, by definition, means that the Jacobian matrix of $\varphi_{t}$,
\begin{equation}
D\varphi_{t}\equiv\frac{\partial\varphi_{t}(\boldsymbol{z})}{\partial\boldsymbol{z}},
\end{equation}
 is symplectic. 

In general, a $2n\times2n$ matrix $S$ is called symplectic if $S^{T}JS=J,$
where 
\begin{equation}
J=\left(\begin{array}{cc}
0 & I_{n\times n}\\
-I_{n\times n} & 0
\end{array}\right),
\end{equation}
which defines an almost complex structure on $\mathbb{R}^{2n},$ i.e.,
$J:\mathbb{R}^{2n}\rightarrow\mathbb{R}^{2n}$ and $J^{2}=-1.$ More
fundamentally, the symplecticity of the solution map $\varphi_{t}$
is a geometric property. The geometric form of the canonical Hamilton's
equation in $\mathbb{R}^{2n}$ is 
\begin{equation}
i_{\left(\dot{\boldsymbol{x}},\dot{\boldsymbol{p}}\right)}\Omega=dH,\label{eq:iOm}
\end{equation}
where 
\begin{equation}
\Omega\equiv-d\gamma=-d\left(\sum_{i=1}^{n}p_{i}dx^{i}\right)\label{eq:Om}
\end{equation}
is the canonical symplectic form. In Eq.\,(\ref{eq:iOm}), ``$i$''
denotes inner product between the 2-form $\Omega$ and the vector
field $\left(\dot{\boldsymbol{x}},\dot{\boldsymbol{p}}\right)$, and
in Eq.\,(\ref{eq:Om}) ``$d$'' denotes exterior derivative. 

The solution map $\varphi_{t}$ being symplectic means geometrically
that $\Omega$ is an invariant of the dynamics, i.e.,
\begin{equation}
\varphi_{t}^{*}\Omega=\Omega,\label{eq:pbOm}
\end{equation}
where $\varphi_{t}^{*}$ denotes the pullback by $\varphi_{t}$. An
immediate consequence of Eq.\,(\ref{eq:pbOm}) is the volume-form
conservation \citep{marsden2013introduction}, 
\begin{align}
\varphi_{t}^{*} & \Lambda^{2n}=\Lambda^{2n},\label{eq:pbvol}
\end{align}
where 
\begin{equation}
\Lambda^{2n}\equiv-\frac{(-1)^{n(n-1)/2}\Omega\wedge\cdots\wedge\Omega}{n!}\text{\,\, (\ensuremath{n} \ensuremath{\text{times)},}}
\end{equation}
is the volume-form defined by the symplectic 2-form $\Omega$ in the
$2n$ dimensional phase space. Equation (\ref{eq:pbvol}) is the well-known
Liouville theorem in the geometric form. Note that the dimension of
$\Omega$, as a 2-form on $\mathbb{R}^{2n},$ is $n(2n-1),$ and the
dimension of $\Lambda^{2n}$ is 1. Volume conservation (\ref{eq:pbvol}),
or incompressibility is necessary but not sufficient for symplecticity
(\ref{eq:pbOm}). 

One may be content with the volume conservation as stated in the classical
Liouville theorem and question the importance of symplecticity. The
fact is that from the viewpoint of physics, phase space volume can
only be meaningfully defined through the symplectic 2-form $\Omega$.
Recall that length, area, and volume in spacetime can only be defined
by the metric tensor that is determined by Einstein's equation. However,
the spacetime metric tensor does not define a volume-form in phase
space, the symplectic 2-form $\Omega$ does \citep{Qin2005Fields,Qin2007Geometric}.
At the fundamental level, phase space volume conservation is one of
many properties of symplectic maps. 

\begin{figure}[ht]
\centering \includegraphics[width=14cm]{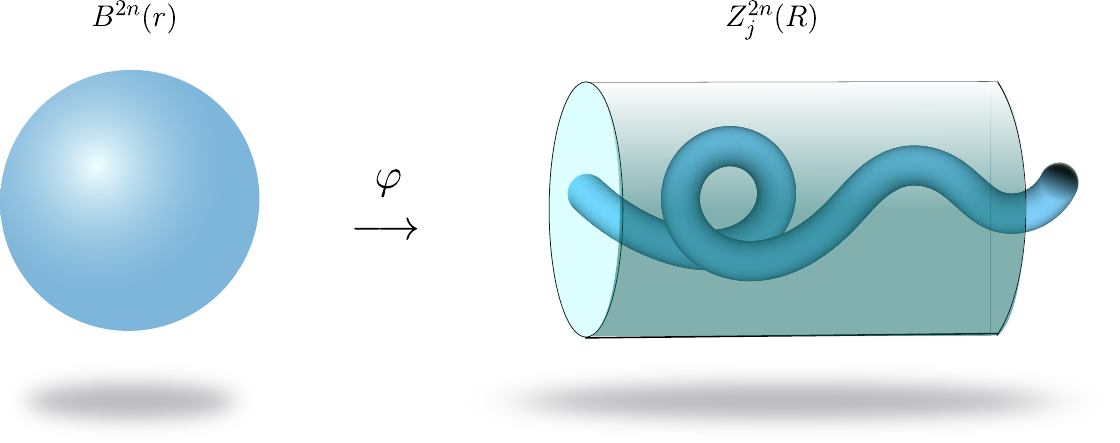} \caption{Gromov's non-squeezing theorem. No symplectic map $\varphi$ can squeeze
the ball $B^{2n}(r)$ into the cylinder $Z_{j}^{2n}(R)$ when $r>R.$
But for any $r\le R,$ the ball $B^{2n}(r)$ is already inside the
cylinder $Z_{j}^{2n}(R)$. The ball $B^{2n}(r)$ in phase space is
``rigid''. }
\label{B2Z}
\end{figure}

In addition to incompressibility (volume-preserving), the non-squeezability
of symplectic maps proved by Gromov in 1985 is the most celebrated
one \citep{Gromov1985,Stewart1987,Hofer1994,deGosson06,Gosson2009,Gosson2009a,Gosson2011,Gosson2013,Gosson2016}.
Gromov's non-squeezing theorem states: If there exists a symplectic
map $\varphi$ in $\mathbb{R}^{2n}$ sending the ball $B^{2n}(r)$
into some cylinder $Z_{j}^{2n}(R),$ then $r\le R.$ (See Fig.\,\ref{B2Z})
Here, 
\begin{align}
B^{2n}(r) & \equiv\left\{ \left(x^{1},x^{2},...,x^{n},p_{1},p_{2},...,p_{n}\right)\left|\sum_{i=1}^{n}\left(p_{i}^{2}+x^{i2}\right)<r^{2}\right.\right\} ,\label{eq:B2n}\\
Z_{j}^{2n}(R) & \equiv\left\{ \left(x^{1},x^{2},...,x^{n},p_{1},p_{2},...,p_{n}\right)\left|p_{j}^{2}+x^{j2}<R^{2}\right.\right\} .\label{eq:Z2n}
\end{align}
This is a strong constraint on symplectic maps considering the fact
that $B^{2n}(r=R)\subset Z_{j}^{2n}(R).$ The ball $B^{2n}(R)$ is
already in the cylinder $Z_{j}^{2n}(R)$, there is no need to squeeze
it. But if we increase the size of the ball by an infinitesimal amount
$\varepsilon$, then we can't push the ball $B^{2n}(R+\varepsilon)$
into the cylinder $Z_{j}^{2n}(R)$ via any symplectic map. 

In $\mathbb{R}^{2n}$, the Euclidean metric $\left\langle \,,\,\right\rangle $,
the canonical symplectic form $\Omega$, and the almost complex structure
$J$ are compatible, i.e.,
\begin{equation}
\left\langle \boldsymbol{a},\boldsymbol{b}\right\rangle =\Omega(\boldsymbol{a},-J\boldsymbol{b}).\label{eq:EuclideanMetric}
\end{equation}
The distance in Eqs.\,(\ref{eq:B2n}) and (\ref{eq:Z2n}) is measured
using this metric. In addition, all maps in the present study are
assumed to be diffeomorphisms (smooth) unless explicitly stated otherwise.

It is easily shown \citep{Gosson2013} that the non-squeezing theorem
is equivalent to the fact that under a symplectic map $\varphi,$
the projection or footprint of $B^{2n}(r)$ on any $(x^{j},p_{j})$
plane is always larger than $\pi r^{2},$ i.e., 
\begin{equation}
\text{Area of }\left\{ \left(x^{j},p_{j}\right)\left|\left(x^{1},x^{2},...,x^{n},p_{1},p_{2},...,p_{n}\right)\in\varphi\left(B^{2n}(r)\right)\right.\right\} \ge\pi r^{2}.
\end{equation}
Non-squeezability stipulates that the phase space footprint on each
$(x^{j},p_{j})$ plane $(1\le j\le n)$ is non-diminishing relative
to an initial set that is a ball $B^{2n}(r)$ in phase space. 

Non-squeezability puts strong constraints on phase space engineering.
Take the example of the Maxwell demon discussed in Sec.\,\ref{sec:pse}.
Starting from a phase space ball in $\mathbb{R}^{6}$, after sending
the system through a Maxwell demon in the $(x,p_{x})$ plane, we may
wish to reduce the footprint of the particle system in the $(x,p_{x})$
plane to reduce the energy cost. This would be achievable by squeezing
a certain amount of phase space volume across the $(y,p_{x})$ or
$(z,p_{z})$ plane if phase space volume conservation were the only
constraint. But, the non-squeezing theorem prohibits this type of
operation. For a given initial distribution $f_{0}$, the corresponding
Gardner ground state may not be accessible by symplectic maps. 

If the initial set $V$ of particles in phase space is not a ball,
what will be the minimum footprint in a conjugate plane $(x^{j},p_{j})$
$(1\le j\le n)$ under a symplectic map $\varphi$? In this case,
we can search for the largest ball that can be squeezed inside $V$
by a symplectic map,
\begin{equation}
C_{0}(V)\equiv\sup_{\varphi\in Sym}\left\{ \pi r^{2}\left|\varphi\left(B^{2n}(r)\right)\subset V\right.\right\} ,\label{eq:C0V}
\end{equation}
where $Sym$ is the set of all symplectic diffeomorphisms (symplectomorphisms)
in $\mathbb{R}^{2n}.$ Obviously, $C_{o}(V)$ is a symplectic invariant
and can be used as a measure of the minimum footprint. Similarly,
we can also search for the smallest cylinder that $V$ can be squeezed
into by a symplectic map, 
\begin{equation}
C_{1}(V)\equiv\inf_{\varphi\in Sym}\left\{ \pi R^{2}\left|\varphi\left(V\right)\subset\right.Z_{j}^{2n}(R)\right\} ,\label{eq:C1V}
\end{equation}
and $C_{1}(V)$ is another symplectic invariant that can be used to
measure the minimum footprint. Because there exist symplectic maps
to switch $(x_{j},p^{j})$ and $(x_{i},p^{i})$ for any $i$ and $j,$
the definition of $C_{1}(V)$ is independent of $j.$ According to
the non-squeezing theorem, 
\[
C_{0}(V)<C_{1}(V).
\]
Thus, for the purpose of phase space engineering, $C_{0}(V)$ and
$C_{1}(V)$ can be used to obtain a good estimate of the minimum footprint
on any phase space conjugate plane $(x^{j},p_{j})$. Here, $C_{0}(V)$
and $C_{1}(V)$ are examples of a family of symplectic invariants
called symplectic capacities \citep{Ekeland1989,Ekeland1990,Hofer1994,Gosson2009,Gosson2016,McDuff2017}
that characterize symplectic maps in a similar manner that phase space
volume characterizes volume-preserving maps. In fact, there are infinitely
many symplectic capacities. It turns out that $C_{0}(V)$ is the smallest
and $C_{1}(V)$ the largest. 

We now display a proof of the non-squeezing theorem for the case of
linear symplectic maps to illustrate the physics related to minimum
footprints and symplectic capacities. The proof is a slightly modified
version of a known proof given in Refs.\,\citep{Gosson2016,McDuff2017}.
(The proof of the general non-squeezing theorem is difficult but not
inaccessible to physicists.) 

Linear symplectic maps are the solutions of linear Hamiltonian systems,
which are found in many branches of physics. In particular, the dynamics
of high-intensity charged particle beams in accelerators, storage
rings, and transport/focusing channels is governed by time-dependent
linear Hamiltonian systems \citep{Davidson01-all,Qin2009PRL-GKV,Qin2010PRL,Qin2011PoP-CSKV,Qin2013PRL2,Qin14-044001,Qin2013PRL-GKVS,Chung16PRL,Chung18,Qin2019LH}.
For non-thermal fusion technologies, the dynamics of high-intensity
particle beams is expected to play an important role. The solution
of a linear Hamiltonian system is a linear map specified by a $2n\times2n$
symplectic matrix $S,$
\begin{equation}
\varphi:\boldsymbol{z}\mapsto\bar{\boldsymbol{z}}=\varphi(\boldsymbol{z})=S\boldsymbol{z}.
\end{equation}
Here, $\boldsymbol{z}=\left(x^{1},x^{2},...,x^{n},p_{1},p_{2},...,p_{n}\right).$ 

For linear symplectic maps, Gromov's non-squeezing theorem reads:
If there exist a symplectic matrix $S$ and an integer $j\le n$ such
that $S\left(B^{2n}(r)\right)\subset Z_{j}^{2n}(R)$, then $R\ge r.$ 

Without losing generality, we prove the case of $r=1$ and $j=1.$
The strategy of proof is to show that for some $\boldsymbol{z}_{0}$
on the boundary of $B^{2n}(1),$ the absolute value of the $x_{1}$
or $p^{1}$ coordinate of $S\boldsymbol{z}_{0}$ is no less than $1.$
However, by definition, $B^{2n}(1)$ is open, and we need to select
a sequence of points in $B^{2n}(1)$ to approach $\boldsymbol{z}_{0}.$
Using the metric defined in Eq.\,(\ref{eq:EuclideanMetric}), we
select the orthonormal bases $(\boldsymbol{a},\boldsymbol{b})$ for
the $(x_{1},p^{1})$ coordinates. The images of $\boldsymbol{a}$
and $\boldsymbol{b}$ under $S^{T}$ are $\bar{\boldsymbol{a}}=S^{T}\boldsymbol{a}$
and $\bar{\boldsymbol{b}}=S^{T}\boldsymbol{b}$. Consider 
\begin{equation}
1=\Omega(\boldsymbol{a},\boldsymbol{b})=\Omega(\bar{\boldsymbol{a}},\bar{\boldsymbol{b}})=\left\langle \bar{\boldsymbol{a}},-J\bar{\boldsymbol{b}}\right\rangle \le\left|\bar{\boldsymbol{a}}\right|\left|J\bar{\boldsymbol{b}}\right|=\left|\bar{\boldsymbol{a}}\right|\left|\bar{\boldsymbol{b}}\right|,
\end{equation}
where the second equal sign is because $S^{T}$ is symplectic, and
inequality is the Cauchy-Schwartz inequality. Therefore, either $\left|\bar{\boldsymbol{a}}\right|\ge1$
or $\left|\bar{\boldsymbol{b}}\right|\ge1$. Assume $\left|\bar{\boldsymbol{a}}\right|\ge1$
and let
\begin{equation}
\boldsymbol{z}_{\varepsilon}=(1-\varepsilon)\frac{\bar{\boldsymbol{a}}}{\left|\bar{\boldsymbol{a}}\right|}\in B^{2n}(1),\quad\epsilon\in(0,1).
\end{equation}
The $x^{1}$ coordinate of $S\boldsymbol{z}_{\varepsilon}$ is
\begin{align}
\left\langle \boldsymbol{a},S\boldsymbol{z}_{\varepsilon}\right\rangle  & =\Omega(\boldsymbol{a},-JS\boldsymbol{z}_{\varepsilon})=\Omega(S^{T}\boldsymbol{a},-S^{T}JS\boldsymbol{z}_{\varepsilon})=\Omega(\bar{\boldsymbol{a}},-J\boldsymbol{z}_{\varepsilon})\nonumber \\
 & =\left\langle \bar{\boldsymbol{a}},(1-\varepsilon)\frac{\bar{\boldsymbol{a}}}{\left|\bar{\boldsymbol{a}}\right|}\right\rangle =(1-\varepsilon)\left|\bar{\boldsymbol{a}}\right|\ge(1-\varepsilon).
\end{align}
If $\left|\bar{\boldsymbol{b}}\right|\ge1$, we select $\boldsymbol{z}_{\varepsilon}=(1-\varepsilon)\bar{\boldsymbol{b}}/\left|\bar{\boldsymbol{b}}\right|\in B^{2n}(1),$
and its $p_{1}$ coordinate will be no less than $1-\varepsilon.$
Thus, the radius of $Z_{j}^{2n}(R)$ should be no less than $1-\varepsilon,$
since $S\boldsymbol{z}_{\varepsilon}\in Z_{j}^{2n}(R).$ Let $\varepsilon\rightarrow0$
and we have $R\ge1.$ This completes the proof. 

In the proof, we find on the boundary of $B^{2n}(1)$ a point $\boldsymbol{z}_{0}$,
whose footprint under any map $S$ on the $(x^{1},p_{1})$ plane sticks
out the unit circle. But neither $\boldsymbol{z}_{0}$ nor $S\boldsymbol{z}_{0}$
is necessarily on the $(x^{1},p_{1})$ plane. $\boldsymbol{z}$ can
be anywhere on the surface of $B^{2n}(1),$ depending on $S.$ The
fact that the footprint of $S\left(B^{2n}(1)\right)$ sticks out the
unit circle on the $(x^{1},p_{1})$ plane for any $S$ is attributed
to the information on the entire $B^{2n}(1)$, and this situation
is exactly the same for every $(x^{j},p_{j})$ plane. In terms of
physics and phase space engineering, the minimum footprint of a group
of particles on any $(x^{j},p_{j})$ plane is the same as measured
by a symplectic capacity, which is a scalar assigned to the entire
group of particles. But we have choices as to which symplectic capacity
to use. 

For the last topic in this section, let's go back to the canonical
Hamilton's equation for charged particles specified by Eqs.\,(\ref{eq:HE})-(\ref{eq:H}).
The kinetic momentum $\boldsymbol{p}$ is the sum of kinetic momentum
$m\dot{\boldsymbol{x}}$ and unit-normalized vector potential $\boldsymbol{A}(\boldsymbol{x},t)$
at the particle's location. If we choose to use $\boldsymbol{x}$
and $\boldsymbol{v}=\dot{\boldsymbol{x}}$ as the phase space coordinates,
Hamilton's equation still assumes the form of Eq.\,(\ref{eq:HE})
except that the canonical symplectic form $\Omega$ will be replaced
by a non-canonical symplectic form $\omega$, i.e., 
\begin{align}
i_{\left(\dot{\boldsymbol{x}},m\dot{\boldsymbol{v}}\right)}\omega & =dH,\label{eq:nonCalHe}\\
\omega & =\sum_{i=1}^{3}dx^{i}\wedge\left(mdv_{i}+\frac{q}{c}dA_{i}(\boldsymbol{x},t)\right),\label{eq:omega}\\
H & =q\phi(\boldsymbol{x},t)+\frac{1}{2}m\boldsymbol{v}^{2}.\label{eq:nonCalH}
\end{align}
 It is straightforward to verify that Eqs.\,(\ref{eq:nonCalHe})-(\ref{eq:nonCalH})
are equivalent to the familiar equations of motion,
\begin{align*}
\dot{\boldsymbol{x}} & =\boldsymbol{v},\\
\dot{\boldsymbol{v}} & =\frac{q}{m}\left(\boldsymbol{E}(\boldsymbol{x},t)+\frac{1}{c}\boldsymbol{v}\times\boldsymbol{B}(\boldsymbol{x},t)\right).
\end{align*}

If the phase space engineering needs to consider the electromagnetic
field self-consistently generated by the fusion plasma itself, the
charged particle-electromagnetic field system is an infinite dimensional
non-canonical Hamiltonian system specified by the Morrison-Marsden-Weinstein
(MMW) bracket \citep{Iwinski1976,Morrison1980Maxwell,Weinstein1981comments,morrison1982AIP,Marsden1982,Marsden1982Hamiltonian},

\begin{align}
\left\{ \mathcal{F},\mathcal{G}\right\}  & (\boldsymbol{E},\boldsymbol{B},f)=\int f\left\{ \frac{\delta\mathcal{F}}{\delta f},\frac{\delta\mathcal{G}}{\delta f}\right\} _{\boldsymbol{xv}}d\boldsymbol{x}d\boldsymbol{v}\nonumber \\
 & +\int\left[\frac{\delta\mathcal{F}}{\delta\boldsymbol{E}}\cdot\left(\triangledown\times\frac{\delta\mathcal{G}}{\delta\boldsymbol{B}}\right)-\frac{\delta\mathcal{G}}{\delta\boldsymbol{E}}\cdot\left(\triangledown\times\frac{\delta\mathcal{F}}{\delta\boldsymbol{B}}\right)\right]d\boldsymbol{x}\nonumber \\
 & +\int\left(\frac{\delta\mathcal{F}}{\delta\boldsymbol{E}}\cdot\frac{\partial f}{\partial\boldsymbol{v}}\frac{\delta\mathcal{G}}{\delta f}-\frac{\delta\mathcal{G}}{\delta\boldsymbol{E}}\cdot\frac{\partial f}{\partial\boldsymbol{v}}\frac{\delta\mathcal{F}}{\delta f}\right)d\boldsymbol{x}d\boldsymbol{v}\nonumber \\
 & +\int f\boldsymbol{B}\cdot\left(\frac{\partial}{\partial\boldsymbol{v}}\frac{\delta\mathbf{\mathrm{\mathcal{F}}}}{\delta f}\times\frac{\partial}{\partial\boldsymbol{v}}\frac{\delta\mathcal{G}}{\delta f}\right)d\boldsymbol{x}d\boldsymbol{v},\label{eq:MMWB}
\end{align}
where $\mathcal{F}$ and $\mathcal{G}$ are functionals of the function
space $\left\{ (\boldsymbol{E},\boldsymbol{B},f)\left|\nabla\cdot\boldsymbol{B}=0,\nabla\cdot\boldsymbol{E}=4\pi q\int fd\boldsymbol{v}\right.\right\} $.
The bracket $\{h,g\}_{\boldsymbol{xv}}$ inside the first term on
the right hand side is the canonical Poisson bracket for functions
$h$ and $g$ of $(\boldsymbol{x},\boldsymbol{v})$,
\begin{equation}
\{h,g\}_{\boldsymbol{xv}}\equiv\frac{\partial h}{\partial\boldsymbol{x}}\frac{\partial g}{\partial\boldsymbol{v}}-\frac{\partial g}{\partial\boldsymbol{x}}\frac{\partial h}{\partial\boldsymbol{v}}.
\end{equation}
Hamilton's equation for the dynamics of a functional $\mathcal{F}$
is 
\begin{equation}
\dot{\mathcal{F}}=\left\{ \mathcal{F},\mathcal{H}\right\} ,\label{eq:Fdot}
\end{equation}
where Hamiltonian functional is
\begin{equation}
\mathcal{H}(f,\boldsymbol{E},\boldsymbol{B})=\frac{m}{2}\int\boldsymbol{v}{}^{2}fd\boldsymbol{x}d\boldsymbol{v}+\frac{1}{8\pi}\int\left(\boldsymbol{E}^{2}+\boldsymbol{B}^{2}\right)d\boldsymbol{x}.\label{eq:CurlyH}
\end{equation}
The equations of motion for the fields are obtained by expressing
them as functionals indexed by $\boldsymbol{x}$ and $\boldsymbol{v}$,
\begin{align}
f\left(\boldsymbol{x},\boldsymbol{v}\right) & =\mathcal{F}\left(\boldsymbol{x},\boldsymbol{v}\right)\equiv\int f\left(\boldsymbol{x}',\boldsymbol{v}'\right)\delta\left(\boldsymbol{x}-\boldsymbol{x}'\right)\delta\left(\boldsymbol{v}-\boldsymbol{v}'\right)d\boldsymbol{x}'d\boldsymbol{v}',\\
\boldsymbol{E}\left(\boldsymbol{x}\right) & =\mathcal{E}\left(\boldsymbol{x}\right)\equiv\int\boldsymbol{E}\left(\boldsymbol{x}'\right)\delta\left(\boldsymbol{x}-\boldsymbol{x}'\right)d\boldsymbol{x}',\\
\boldsymbol{B}\left(\boldsymbol{x}\right) & =\mathcal{B}\left(\boldsymbol{x}\right)\equiv\int\boldsymbol{B}\left(\boldsymbol{x}'\right)\delta\left(\boldsymbol{x}-\boldsymbol{x}'\right)d\boldsymbol{x}'.
\end{align}
 The Hamiton's equation (\ref{eq:Fdot}) for $\mathcal{F}\left(\boldsymbol{x},\boldsymbol{v}\right)$,
$\mathcal{E}\left(\boldsymbol{x}\right),$ and $\mathcal{B}\left(\boldsymbol{x}\right)$
recovers the standard Vlasov-Maxwell equations, 
\begin{align}
 & \frac{\partial f}{\partial t}=-\boldsymbol{v}\cdot\frac{\partial f}{\partial\boldsymbol{x}}-\frac{q}{m}\left(\boldsymbol{E}+\frac{\boldsymbol{v}}{c}\times\boldsymbol{B}\right)\cdot\frac{\partial f}{\partial\boldsymbol{v}},\\
 & \frac{\partial\boldsymbol{E}}{\partial t}=c\nabla\times\boldsymbol{B}-4\pi\int qf\boldsymbol{v}d^{3}\boldsymbol{v}\\
 & \frac{\partial\boldsymbol{B}}{\partial t}=-c\nabla\times\boldsymbol{E}.
\end{align}

\section{Structure-preserving geometric algorithms for phase space engineering\label{sec:spa}}

While phase space incompressibility has been well-known for centuries,
phase space non-squeezability has only been established for 40 years.
The associated dynamic invariants, symplectic capacities, have not
been widely recognized in most physics applications. For example,
Gardner's ground state analysis for plasmas only considered the phase
space incompressibility. When the non-squeezability is included in
the analysis, the ground state is expected to be very different, and
the ground state energy will increase. From the perspective of fusion
energy production, this means less energy can be extracted via conservative
electromagnetic channels. In this sense and at this level of sophistication,
phase space engineering for non-thermal advanced fuel fusion is abstractly
symplectic topology. Admittedly, it is difficult to directly incorporate
phase space non-squeezability and other constraints of symplectic
topology in analytical studies or numerical simulations of phase space
engineering. For instance, the minimum footprint estimates according
to Eqs.\,(\ref{eq:C0V}) and (\ref{eq:C1V}) are important design
parameters, but it is not clear how they can be accurately calculated. 

Fortunately, recent investigations in structure-preserving geometric
algorithms for charged particle-electromagnetic field systems have
offered a solution. The newly discovered structure-preserving geometric
algorithms are able to preserve the non-canonical symplectic structures
given by Eqs.\,(\ref{eq:omega}) and (\ref{eq:MMWB}) exactly in
discretized spacetime settings. Consequently, incompressibility, non-squeezability,
and symplectic capacities are preserved exactly by numerical solutions.
Specifically, arbitrary high-order, explicitly solvable, non-canonical
symplectic algorithms have been found for both charged particle dynamics
governed by Eqs.\,(\ref{eq:nonCalHe})-(\ref{eq:nonCalH}) \citep{Qin2008VI-PRL,he2015Hamiltonian,xiao2015explicit,he2017explicit,zhang2016explicit,zhou2017explicit,Xiao2019Relativisitic,Xiao2021Explicit}
and the Vlasov-Maxwell systems governed by Eqs.\,(\ref{eq:MMWB})-(\ref{eq:CurlyH})
\citep{Squire4748,squire2012geometric,xiao2013variational,xiao2015explicit,xiao2015variational,he2015Hamiltonian,he2016hamiltonian,qin2016canonical,xiao2017local,kraus2017gempic,Morrison2017,burby2017finite,Xiao2018review,Xiao2019field,Xiao2021Explicit,Glasser2020,Wang2021,Kormann2021,Perse2021,Glasser2022,CamposPinto2022,Burby2023}.
With these structure-preserving geometric algorithms, we can numerically
calculate the symplectic capacities defined in Eqs.\,(\ref{eq:C0V})
and (\ref{eq:C1V}) as the minimum footprints for phase space engineering
designs. 

In this section, we explain in detail the mechanisms by which these
algorithms preserve exactly the symplecticity and thus incompressibility,
non-squeezability, and symplectic capacities of the Hamiltonian systems.
To be concise, we will focus only on the high-order, explicitly solvable,
non-canonical symplectic algorithms for the charged particle dynamics
governed by Eqs.\,(\ref{eq:nonCalHe})-(\ref{eq:nonCalH}). 

Let $\varphi_{\Delta t}$ denote the exact solution map of Eq.\,(\ref{eq:nonCalHe})
in a time interval $\Delta t$,
\begin{equation}
\varphi_{\Delta t}:\boldsymbol{z}(t)\mapsto\boldsymbol{z}(t+\Delta t)=\varphi_{\Delta t}(\boldsymbol{z}).
\end{equation}
 $\varphi_{\Delta t}$ is symplectic, i.e, 
\begin{equation}
\varphi_{\Delta t}^{*}\omega=\omega.
\end{equation}
Denote by $\nu_{\Delta t}$ a numerical solution map of Eq.\,(\ref{eq:nonCalHe})
in a time interval $\Delta t$,
\begin{equation}
\nu_{\Delta t}:\boldsymbol{z}(t)\mapsto\widetilde{\boldsymbol{z}}(t+\Delta t)=\nu_{\Delta t}(\boldsymbol{z}).\label{eq:nudt}
\end{equation}
$\nu_{\Delta t},$ called one-step map of the algorithm, is an approximation
to $\varphi_{\Delta t}$. We would like $\nu_{\Delta t}$ to be exactly
symplectic as well, i.e., 
\begin{equation}
\nu_{\Delta t}^{*}\omega=\omega.\label{eq:nvom}
\end{equation}
Even though $\nu_{\Delta t}$ is not exactly the same as $\varphi_{\Delta t}$
in general, we prefer $\nu_{\Delta t}$ to come from the same family
of symplectomorphisms because all important properties of Hamiltonian
dynamics, such as incompressibility and non-squeezability, are associated
with the symplectomorphisms as a group, instead of with a specific
member of the group. An algorithm is called symplectic if its one-step
map is symplectic. 

For a general non-canonical symplectic system, condition (\ref{eq:nvom})
is difficult to satisfy unless $\nu_{\Delta t}$ is the exact solution
$\varphi_{\Delta t}$. In fact, no generic symplectic algorithm exists
for general non-canonical systems. But for a specific given non-canonical
system, bespoke symplectic algorithms might exist. Equations\,(\ref{eq:nonCalHe})-(\ref{eq:nonCalH})
for charged particle dynamics are such a system. The bespoke symplectic
algorithms are constructed by a Hamiltonian splitting technique.

The symplectic condition (\ref{eq:nvom}) is satisfied if $\nu_{\Delta t}$
is an exact solution. Of course, we can't write down the exact solution
corresponding to the $\omega$ defined by Eq.\,(\ref{eq:omega})
and $H$ defined by Eq.\,(\ref{eq:nonCalH}). However, observe that
for the $\omega$ defined by Eq.\,(\ref{eq:omega}), the symplectic
condition (\ref{eq:nvom}) is satisfied exactly by the exact solution
$\varphi_{\Delta t}$ generated by any Hamiltonian function. For many
simple Hamiltonian functions, the exact solution maps can be written
down explicitly, and the compositions of these exact solution maps
will also preserve the $\omega$ defined by Eq.\,(\ref{eq:omega}). 

If we can split the Hamiltonian $H$ defined by Eq.\,(\ref{eq:nonCalH})
into different parts such that the exact solution for each sub-Hamiltonian
system can be written down explicitly, then these exact solutions
of the sub-systems can be composed to approximate the exact solution
of $H$ to arbitrary high orders, and such approximate solutions automatically
satisfy the symplectic condition (\ref{eq:nvom}). Such a splitting
was found. Specifically, we split $H$ into four sub-Hamiltonians
\citep{he2015Hamiltonian,xiao2015explicit,he2017explicit,Xiao2019Relativisitic,Xiao2021Explicit}
,
\begin{align}
H & =H_{\phi}+H_{x}+H_{y}+H_{z},\\
H_{\phi} & =q\phi(\boldsymbol{x},t),\\
H_{x} & =\frac{1}{2}mv_{x}^{2},\,\,H_{y}=\frac{1}{2}mv_{y}^{2},\,\,H_{z}=\frac{1}{2}mv_{z}^{2}.
\end{align}
Here, $(x,y,z)$ are the Cartesian coordinates of the configuration
space, and $(v_{x},v_{y},v_{z})$ are the corresponding velocity coordinates. 

For the sub-system specified by $H_{\phi},$ Eq.\,(\ref{eq:nonCalHe})
is 
\begin{equation}
i_{\left(\dot{\boldsymbol{x}},m\dot{\boldsymbol{v}}\right)}\omega=q\nabla\phi(\boldsymbol{x},t),\label{eq:HphiE}
\end{equation}
which, in terms of $\dot{\boldsymbol{x}}$ and $\dot{\boldsymbol{v}}$,
is 
\begin{align}
\dot{\boldsymbol{x}} & =0,\\
\dot{\boldsymbol{v}} & =\frac{q}{m}\boldsymbol{E}(\boldsymbol{x},t),
\end{align}
where 
\begin{equation}
\boldsymbol{E}(\boldsymbol{x},t)=-\nabla\phi(\boldsymbol{x},t)-\frac{\partial\boldsymbol{A}(\boldsymbol{x},t)}{\partial t}.\label{eq:E}
\end{equation}
Note that in Eq.\,(\ref{eq:E}), the time derivative of the vector
potential enters through the non-canonical symplectic form $\omega$
defined in Eq.\,(\ref{eq:omega}). The exact solution map of Eq.\,(\ref{eq:HphiE})
can be explicitly written down, 
\begin{equation}
\varphi_{\phi,\Delta t}:\begin{cases}
\boldsymbol{x}(t+\Delta t)=\boldsymbol{x}(t),\\
\boldsymbol{v}(t+\Delta t)=\boldsymbol{v}(t)+\frac{q}{m}\int_{0}^{\Delta t}\boldsymbol{E}\left(\boldsymbol{x}(t),t+\tau\right)d\tau.
\end{cases}
\end{equation}
For the sub-system specified by $H_{x}$, Eq.\,(\ref{eq:nonCalHe})
is 
\begin{equation}
i_{\left(\dot{\boldsymbol{x}},m\dot{\boldsymbol{v}}\right)}\omega=mv_{x}dv_{x},\label{eq:Hphix}
\end{equation}
which is
\begin{gather}
\dot{x}=v_{x},\,\,\dot{y}=0,\,\,\dot{z}=0,\label{eq:xyz}\\
\dot{v_{x}}=0,\,\,\dot{v_{y}}=-\frac{q}{m}B_{z}(\boldsymbol{x},t)v_{x},\,\,\dot{v_{z}}=\frac{q}{m}B_{y}(\boldsymbol{x},t)v_{x}.\label{eq:v}
\end{gather}
In Eq.\,(\ref{eq:v}), the magnetic field $\boldsymbol{B}(\boldsymbol{x},t)=\nabla\times\boldsymbol{A}(\boldsymbol{x},t)$
also enters through the non-canonical symplectic form $\omega$. Equations
(\ref{eq:xyz}) and (\ref{eq:v}) are more complicated than those
of $H_{\phi}.$ But the exact solution map of $H_{x}$ can also be
written down explicitly \citep{he2015Hamiltonian,xiao2015explicit,he2017explicit,Xiao2019Relativisitic,Xiao2021Explicit},

\begin{equation}
\varphi_{x,\Delta t}:\begin{cases}
x(t+\Delta t)=x(t)+v_{x}(t)\Delta t,\\
y(t+\Delta t)=y(t),\\
z(t+\Delta t)=z(t),\\
v_{x}(t+\Delta t)=v_{x}(t)\\
v_{y}(t+\Delta t)=v_{y}(t)-\frac{q}{m}\int_{0}^{\Delta t}v_{x}(t)B_{z}\left(x(t)+\tau v_{x}(t),y(t),z(t),t+\tau\right)d\tau,\\
v_{z}(t+\Delta t)=v_{z}(t)+\frac{q}{m}\int_{0}^{\Delta t}v_{x}(t)B_{y}\left(x(t)+\tau v_{x}(t),y(t),z(t),t+\tau\right)d\tau.
\end{cases}
\end{equation}
The exact solution maps $\varphi_{y,\Delta t}$ and $\varphi_{z,\Delta t}$
for the sub-systems $H_{y}$ and $H_{z}$ can be explicitly written
down in a similar manner. 

It is crucial to recognize that the exact solution maps $\varphi_{x,\Delta t}$
$\varphi_{y,\Delta t},$ and $\varphi_{z,\Delta t}$ are written down
in the Cartesian coordinates. It has been shown that for an orthogonal,
uniform curvilinear coordinate system $(q^{1},q^{2},q^{3})$ in $\mathbb{R}^{3}$,
the exact solution for sub-system $H_{q^{i}}$ can also be written
down explicitly \citep{Xiao2021Explicit}. An orthogonal coordinate
system is called uniform if 
\begin{equation}
\frac{\partial h_{i}}{\partial q^{i}}=0,\,(i=1,2,3),
\end{equation}
where $h_{i}$ $(i=1,2,3)$ are the Lamé coefficients. If a curvilinear
coordinate system is not orthogonal or non-uniformly orthogonal, the
solution maps for the subsystem defined by $H_{q^{i}}$ can't be written
down explicitly in general \citep{Xiao2021Explicit}. 

Arbitrary high-order symplectic algorithms for the original Hamiltonian
system (\ref{eq:nonCalHe})-(\ref{eq:nonCalH}) can be constructed
from different compositions of $\varphi_{\phi,\Delta t}$, $\varphi_{x,\Delta t}$
$\varphi_{y,\Delta t},$ and $\varphi_{z,\Delta t}$. Two of the 1st-order
algorithms for $H$ are
\begin{align}
\nu_{\Delta t}^{(1)} & =\varphi_{\phi,\Delta t}\circ\varphi_{x,\Delta t}\circ\varphi_{y,\Delta t}\circ\varphi_{z,\Delta t},\\
\nu_{\Delta t}^{*(1)} & =\varphi_{z,\Delta t}\circ\varphi_{y,\Delta t}\circ\varphi_{x,\Delta t}\circ\varphi_{\phi,\Delta t}.
\end{align}
A 2nd-order algorithm for $H$ can be built as
\begin{equation}
\nu_{\Delta t}^{(2)}=\nu_{\Delta t/2}^{(1)}\circ\nu_{\Delta t/2}^{*(1)},
\end{equation}
which is the familiar Strang splitting. From a given $2l$th-order
algorithm $\nu_{\Delta t}^{(2l)},$ a $2(l+1)$th-order algorithm
can be constructed as follows,
\begin{align}
\nu_{\Delta t}^{(2l+2)} & =\nu_{\alpha_{l}\Delta t}^{(2l)}\circ\nu_{\beta_{l}\Delta t}^{(2l)}\circ\nu_{\alpha_{l}\Delta t}^{(2l)},\\
\alpha_{l} & \equiv\left(2-\frac{1}{2l+1}\right)^{-1},\thinspace\thinspace\beta_{l}\equiv1-2\alpha_{l}.
\end{align}

As emphasized above, all these algorithms preserve exactly the symplectic
structure $\omega$ defined in Eq.\,(\ref{eq:omega}) as the exact
solutions do. 

In the stability analysis of algorithms when applied to linear differential
ODEs with constant coefficients, algorithms are often called implicit
if the derivative evaluation depends on the values of dependent variables
at future time-steps. If this definition is strictly followed, the
symplectic algorithms described above are implicit. However, there
is no need for numerical root-finding as in standard implicit algorithms.
To be precise, the algorithms developed here for charged particle
dynamics are explicitly solvable, implicit, and symplectic. 

These non-canonical symplectic algorithms can be used as upgrades
of volume-preserving but non-symplectic algorithms \citep{he2015volume,zhang2015volume,He16Higher,Zhang2016,He2016HigherRela,Tu2016,Matsuyama2017,Higuera2017}
for charged particle dynamics, including the widely adopted Boris
algorithm \citep{Boris70,Stoltz02,Penn03,Qin2013Boris,Winkel2015,Winkel2015b,Hairer2018,ellison2015comment,Umeda2018,Zenitani2018,Hairer2020,Zenitani2020,Xiao2020Slow,Fu2022,Hairer2022,Lubich2023,Hairer2023}.

Finally, these non-canonical symplectic algorithms can be further
upgraded to become structure-preserving geometric particle-in-cell
algorithms for the Vlasov-Maxwell system when the effects of self-consistently
generated electromagnetic fields are important \citep{Squire4748,squire2012geometric,xiao2013variational,xiao2015explicit,xiao2015variational,he2015Hamiltonian,he2016hamiltonian,qin2016canonical,xiao2017local,kraus2017gempic,Morrison2017,burby2017finite,Xiao2018review,Xiao2019field,Xiao2021Explicit,Glasser2020,Wang2021,Kormann2021,Perse2021,Glasser2022,CamposPinto2022}.
These structure-preserving algorithms are suitable tools for studying
the physical processes of phase space engineering in non-thermal advanced
fuel fusion, such as electromagnetic energy extraction and phase space
Maxwell demons. 

\section{Conclusions}

The currently envisioned main pathway to fusion energy is the D-T
thermal fusion via magnetic confinement or inertial confinement. Among
all possible light-ion fusion reactions that are potentially suitable
for energy production. D-T fusion is the only one that can be sustained
in a thermalized plasma because of the relatively low temperature
required. The disadvantages are also prominent. The 14MeV fusion neutron
demands new technologies for energy conversion and hardware protection,
and the fuel self-sustainability of tritium imposes a stringent constraint
on the recycling and recovery rate. Advanced fuel fusion using D-He3
or P-B11 might be a viable alternative for commercial fusion energy
production. However, to be practical, advanced fuel fusion needs to
be in non-thermal regimes to avoid excessive radiation loss of energy
because the fusion cross-sections for advanced fuel fusion peak at
much higher energies. This leads to large recirculating power in the
system. Advanced fuel fusion trades the requirement of a large of
amount recirculating tritium for that of large recirculating power.
To meet the challenge of maintaining non-thermal particle distributions
at a reasonable cost, phase space engineering technologies utilizing
externally injected electromagnetic fields can be applied. For charged
particles in ionized plasmas, injecting high-frequency electromagnetic
fields seems to be the only way to manipulate phase space distributions
faster than the collisional thermalization, and such methods have
been extensively investigated for achieving hot ion modes via $\alpha$-channeling
in D-T fusion. 

In the present study, we have studied the physical process of phase
space engineering, such as the Maxwell demon and electromagnetic energy
extraction, from a theoretical and algorithmic perspective. We demonstrated
that the operational space of phase space engineering is limited by
the underpinning symplectic dynamics of charged particles. Volume
conservation, or incompressibility, according to the well-known Liouville
theorem is just one of many phase space constraints. Gromov's non-squeezing
theorem determines the minimum footprint of the charged particles
on every conjugate phase space plane. Other constraints, such as symplectic
capacities, need to be included in the phase space engineering designs
as well. In this sense and level of sophistication, phase space engineering
is abstractly symplectic topology. To calculate the minimum footprint
of charged particles for phase space engineering and to accurately
simulate the processes of phase space engineering, recently developed
structure-preserving geometric algorithms can be used. The family
of algorithms conserves exactly, on discretized spacetime, symplecticity
and thus incompressibility, non-squeezability, and symplectic capacities.
The algorithms are implicit but explicitly solvable and apply to finite
dimensional non-canonical symplectic dynamics of charged particles
under the influence of external electromagnetic fields, as well as
to the infinite dimensional non-canonical symplectic charged particle-electromagnetic
field system described by the Vlasov-Maxwell equations in the geometric
form. 
\begin{acknowledgments}
This research was supported by the U.S. Department of Energy (DE-AC02-09CH11466).
I thank Prof. N. J. Fisch, Prof. I. Y. Dodin, Prof. A. H. Reiman,
Dr. E. J. Kolmes, Dr. A. S. Glasser, Dr. I. E. Ochs, T. Rubin, Dr.
M. E. Mlodik, Dr. W. M. Nevins, Dr. W. W. Lee, and Dr. M. A. de Gosson
for fruitful discussions. The present study is inspired by their groundbreaking
contributions. 
\end{acknowledgments}

\bibliographystyle{apsrev4-2}
\bibliography{../Refs/Qin,../Refs/Refs}

\end{document}